\documentclass[RNAAS]{aastex62}

\usepackage{upgreek}

\begin{document}

\title{Simulated Direct Imaging Detection of Water Vapor for Exo-Earths}

\correspondingauthor{Anna S. Ross}
\email{asr99@nau.edu}

\author[0000-0002-2194-4268]{Anna Sage Ross \includegraphics[scale=0.05]{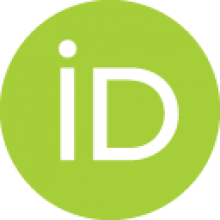}}
\affiliation{
Department of Astronomy and Planetary Science, Northern Arizona University, P.O. Box 6010, Flagstaff, AZ 86011, USA}

\author[0000-0002-3196-414X]{Tyler D. Robinson\includegraphics[scale=0.05]{Orcid-ID.png}}
\affiliation{
Department of Astronomy and Planetary Science, Northern Arizona University, P.O. Box 6010, Flagstaff, AZ 86011, USA}
\altaffiliation{}


\section{}

Habitable planets are 
often defined as terrestrial worlds capable of maintaining surface liquid water  \citep{kastingetal1993}.  As a result, atmospheric water vapor can be a critical indicator of habitability \citep[see review by][]{robinson2018}.  Thus, habitability-themed exoplanet 
investigations emphasize detection of water vapor signatures for their targets.

A variety of concept missions for exoplanet direct imaging in reflected light have seen recent study 
\citep[e.g.,][]{seageretal2015}.  Two such concepts emphasize broad capabilities to detect biosignatures and habitability indicators from Earth-analog exoplanets: the Habitable Exoplanet Observatory \citep[HabEx;][]{gaudietal2018} and the Large UV/Optical/IR (LUVOIR) Surveyor \citep{robergeetal2018}.  
However, the amount 
of water vapor present in habitable exoplanet atmospheres can be diverse
\citep[e.g.,][]{abeletal2011}.  It is thus important to understand how direct imaging in reflected light --- coupled with moderate-resolution spectroscopy --- could be used to detect various water vapor amounts in Earth-like exoplanetary atmospheres.

To investigate water vapor detection for exoplanets, we generated reflectance spectra over a grid of water vapor column masses (given by the integral of a species' mass density over the atmospheric column, yielding the mass of that species per unit area).  As the depth of a gas spectral feature is, to a large degree, controlled by the absorption optical depth across the feature, varying the column mass is akin to varying feature strength.  Our column mass grid spans 10$^{-5}$ to 10$^{1}$\,g\,cm$^{-2}$, where the total gas and water vapor column masses for Earth are $1.0\times10^3$\,g\,cm$^{-2}$ and 3\,g\,cm$^{-2}$, respectively.  We adopt 1~bar of total atmospheric pressure, assume molecular nitrogen is the background atmospheric gas, use an Earth-like surface gravity of 10\,m\,s$^{-2}$, and assume a gray surface albedo of 0.3.  Our calculations generally apply to scenarios where the water vapor column is measured above some opaque ``surface'' (e.g., a solid surface or planet-wide cloud deck).  Quadrature-phase spectra were modeled with the widely-used Spectral Mapping Atmospheric Radiative Transfer (SMART) model \citep[developed by D.~Crisp;][]{meadows&crisp1996}, which is a line-by-line, multiple-scattering radiative transfer tool.

\begin{figure}[]
\begin{center}
\includegraphics[]{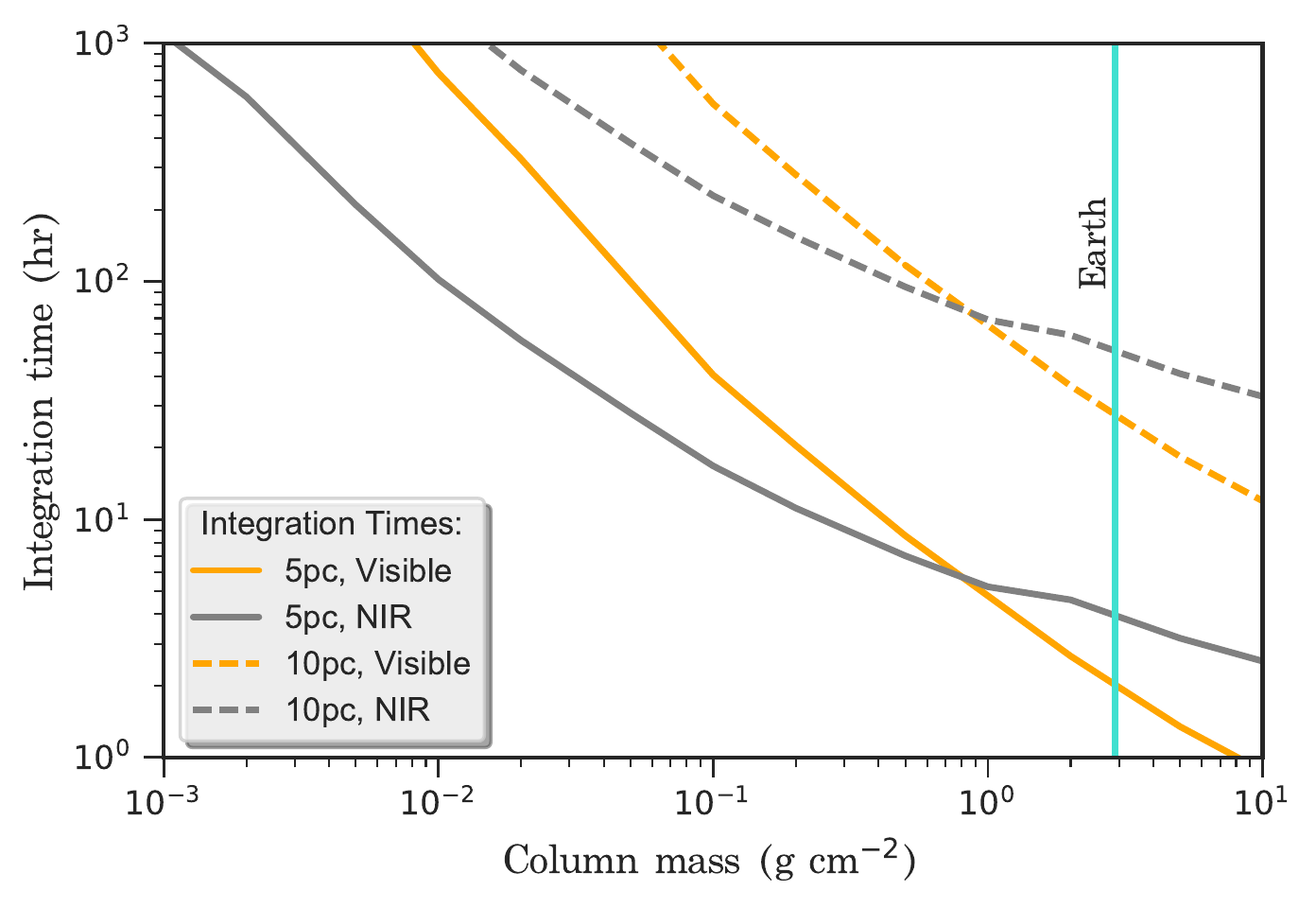}
\caption{Requisite integration times for water vapor detection (at $5\sigma$) over a range of atmospheric water vapor column masses.  Simulation details are described in the text.}
\label{fig:inttimes}
\end{center}
\end{figure}

Requisite integration times required for water vapor detection (at 5$\sigma$), derived from our spectral grid using a public high contrast imaging instrument/noise model \citep{robinsonetal2016}, are shown in Figure~\ref{fig:inttimes}. These calculations are based on a concept with a 4\,m primary mirror, a raw contrast of 10$^{-10}$, and resolving powers ($\lambda/\Delta \lambda$) of 140 and 40 in the visible (0.45--0.95\,$\upmu$m) and near-infrared (NIR; 0.95--1.8\,$\upmu$m), respectively.  Adopting a Sun-like host and using a representative exozodiacal scattered light level of 4.5~``exozodis'' \citep{mennessonetal2019} implies our simulation noise is strongly dominated by exozodiacal light.  Thus, our requisite integration times scale with telescope primary mirror diameter, $D$, according to $D^{-4}$.  Integration time curves are shown for visible- and NIR-only cases for targets at 5 and 10\,pc.

Figure~\ref{fig:inttimes} highlights the competing effects of stronger water vapor bands, coarser spectroscopy, and a lower overall stellar photon flux at longer wavelengths versus weaker water vapor bands, higher-resolution spectroscopy, and a larger overall stellar photon flux at shorter wavelengths.  For targets at both 5 and 10\,pc, the stronger water vapor bands and coarser resolution of the NIR results in shorter requisite integration times for detection than in the visible for column masses smaller than about 1\,g\,cm$^{-2}$.  For larger column masses, otherwise weak visible-wavelength water vapor features become strong enough to enable 
efficient detection of water at visible wavelengths.  A slight flattening of the NIR curves near column masses of 1--2\,g\,cm$^{-2}$ 
is caused by saturation of the 1.4\,$\upmu$m water vapor band.  Critically, water vapor detection times remain below 100~hr for nearby targets with even 1\% the water vapor column mass of Earth.  Water vapor detection for more distant targets can be achieved with integration times spanning days to weeks for water vapor column masses spanning 1--10\% that of Earth.

Lower-resolution NIR spectroscopy is generally optimal for detecting water vapor in the atmospheres of Earth-like exoplanets when using direct imaging in reflected light. This holds true for dry or cold terrestrial planets, whose atmospheres would contain relatively little water vapor. Atmospheres richer in water vapor, such as planets undergoing a moist or runaway greenhouse, could have water vapor detected at visible wavelengths. Understanding details like an exoplanet's size, temperature, and location relative to the habitable zone can aid in determining appropriate wavelength ranges for atmospheric characterization. Overall, water vapor detection for Earth-like exoplanets is quite feasible for future 
direct imaging missions.

\acknowledgments
ASR acknowledges support from the Hooper Undergraduate Research Award at Northern Arizona University. TDR acknowledges support from NASA's Exoplanets Research Program (No.~80NSSC18K0349) and Exobiology Program (No.~80NSSC19K0473), as well as the Nexus for Exoplanet System Science Virtual Planetary Laboratory (No.~80NSSC18K0829).


\begin{thebibliography}{}
\expandafter\ifx\csname natexlab\endcsname\relax\def\natexlab#1{#1}\fi
\providecommand{\url}[1]{\href{#1}{#1}}
\providecommand{\dodoi}[1]{doi:~\href{http://doi.org/#1}{\nolinkurl{#1}}}
\providecommand{\doeprint}[1]{\href{http://ascl.net/#1}{\nolinkurl{http://ascl.net/#1}}}
\providecommand{\doarXiv}[1]{\href{https://arxiv.org/abs/#1}{\nolinkurl{https://arxiv.org/abs/#1}}}

\bibitem[{{Abel} {et~al.}(2011){Abel}, {Frommhold}, {Li}, \&
  {Hunt}}]{abeletal2011}
{Abel}, M., {Frommhold}, L., {Li}, X., \& {Hunt}, K.~L.~C. 2011, Journal of
  Physical Chemistry A, 115, 6805

\bibitem[{{Gaudi} {et~al.}(2018){Gaudi}, {Seager}, {Mennesson}, {Kiessling},
  {Warfield}, {Habitable Exoplanet Observatory Science}, \& {Technology
  Definition Team}}]{gaudietal2018}
{Gaudi}, B.~S., {Seager}, S., {Mennesson}, B., {et~al.} 2018, Nature Astronomy,
  2, 600, \dodoi{10.1038/s41550-018-0549-2}

\bibitem[{{Kasting} {et~al.}(1993){Kasting}, {Whitmire}, \&
  {Reynolds}}]{kastingetal1993}
{Kasting}, J.~F., {Whitmire}, D.~P., \& {Reynolds}, R.~T. 1993, Icarus, 101,
  108

\bibitem[{{Meadows} \& {Crisp}(1996)}]{meadows&crisp1996}
{Meadows}, V.~S., \& {Crisp}, D. 1996, \jgr, 101, 4595

\bibitem[{Mennesson {et~al.}(2019)Mennesson, Kennedy, Ertel, Wyatt, Defrere,
  Debes, Stark, Kasdin, Macintosh, Hinz, {et~al.}}]{mennessonetal2019}
Mennesson, B., Kennedy, G., Ertel, S., {et~al.} 2019, Bulletin of the American
  Astronomical Society, 51, 324

\bibitem[{{Roberge} \& {Moustakas}(2018)}]{robergeetal2018}
{Roberge}, A., \& {Moustakas}, L.~A. 2018, Nature Astronomy, 2, 605,
  \dodoi{10.1038/s41550-018-0543-8}

\bibitem[{{Robinson}(2018)}]{robinson2018}
{Robinson}, T.~D. 2018, in Handbook of Exoplanets, ed. H.~J. {Deeg} \& J.~A.
  {Belmonte} (Springer), 67, \dodoi{10.1007/978-3-319-55333-7_67}

\bibitem[{{Robinson} {et~al.}(2016){Robinson}, {Stapelfeldt}, \&
  {Marley}}]{robinsonetal2016}
{Robinson}, T.~D., {Stapelfeldt}, K.~R., \& {Marley}, M.~S. 2016, \pasp, 128,
  025003, \dodoi{10.1088/1538-3873/128/960/025003}

\bibitem[{{Seager} {et~al.}(2015){Seager}, {Turnbull}, {Sparks}, {Thomson},
  {Shaklan}, {Roberge}, {Kuchner}, {Kasdin}, {Domagal-Goldman}, {Cash},
  {Warfield}, {Lisman}, {Scharf}, {Webb}, {Trabert}, {Martin}, {Cady}, \&
  {Heneghan}}]{seageretal2015}
{Seager}, S., {Turnbull}, M., {Sparks}, W., {et~al.} 2015, in \procspie, Vol.
  9605, Techniques and Instrumentation for Detection of Exoplanets VII, 96050W,
  \dodoi{10.1117/12.2190378}

\end{thebibliography}

\end{document}